\documentclass[12pt]{article}
\usepackage{amsmath, amssymb,amsthm,graphicx,a4wide,hyperref}

\newcommand{\RR}{\mathbb{R}}
\newcommand{\CC}{\mathbb{C}}
\newcommand{\DD}{\mathbb{D}}
\newcommand{\KK}{\mathbb{K}}

\newcommand{\PP}{\mathbb{P}}

\newcommand{\ba}{{\mathbf{a}}}
\newcommand{\bD}{{\mathbf{D}}}
\newcommand{\bH}{{\mathbf{H}}}
\newcommand{\bI}{{\mathbf{I}}}
\newcommand{\bn}{{\mathbf{n}}}
\newcommand{\bU}{{\mathbf{U}}}
\newcommand{\Sx}{\mathbf{\sigma}_x}
\newcommand{\Sy}{\mathbf{\sigma}_y}
\newcommand{\Sz}{\mathbf{\sigma}_z}
\newcommand{\Sp}{\mathbf{\sigma}_+}
\newcommand{\Sm}{\mathbf{\sigma}_-}
\newcommand{\bP}{{\mathbf{P }}}
\newcommand{\bQ}{{\mathbf{Q }}}
\newcommand{\bM}{{\mathbf{M }}}
\newcommand{\bL}{{\mathbf{L }}}
\newcommand{\bV}{{\mathbf{V }}}
\newcommand{\btM}{\widetilde{\mathbf{M }}}
\newcommand{\btL}{\widetilde{\mathbf{L }}}
\newcommand{\bO}{{\mathbf{O }}}
\newcommand{\bS}{{\mathbf{S }}}

\newcommand{\ket}[1]{\left|#1\right\rangle}
\newcommand{\bra}[1]{\left\langle #1\right|}
\newcommand{\bket}[1]{\left\langle #1 \right\rangle}
\newcommand{\braket}[2]{\left\langle #1\right|\!\left.#2 \right\rangle}
\newcommand{\ketbra}[2]{\left|#1 \right\rangle\!\left\langle #2 \right|}
\newcommand{\EE}[2]{\mathbb{E}\left(#1 ~\Big|~ #2 \right)}
\newcommand{\dotex}{\frac{d}{dt}}

\newcommand{\Tr}[1]{{\rm{Tr}}\left(#1\right)}

\title{A tutorial introduction to quantum stochastic master equations
based on the qubit/photon  system}

\author{Pierre Rouchon\thanks{Laboratoire  de Physique de l’Ecole Normale Sup\'{e}rieure, Mines Paris-PSL, Inria,  ENS-PSL, Universit\'{e} PSL, CNRS,   Paris, France. {\tt pierre.rouchon@minesparis.psl.eu} }}

\begin{document}

\maketitle

\begin{abstract}
From the  key composite quantum  system made of a  two-level   system (qubit) and a  harmonic oscillator (photon) with  resonant or  dispersive interactions,  one derives   the corresponding quantum  Stochastic Master Equations (SME) when either the qubits or the photons are measured.  Starting with an elementary  discrete-time formulation based on  explicit formulae for the  interaction propagators, one shows how to include measurement imperfections  and decoherence.  This qubit/photon quantum system  illustrates the Kraus-map structure of general discrete-time SME governing the dynamics of an open quantum system subject to measurement back-action and decoherence induced by the environment. Then, on the qubit/photon system, one explains the passage to a continuous-time mathematical model where the measurement signal is either a  continuous real-value  signal  (typically homodyne or heterodyne signal) or a  discontinuous and integer-value signal  obtained from a counter.  During this derivation, the Kraus map formulation  is  preserved in an infinitesimal way. Such a derivation provides also  an equivalent Kraus-map formulation to the continuous-time  SME  usually expressed  as  stochastic differential equations  driven either by Wiener or Poisson processes.  From such Kraus-map formulation, simple linear numerical integration schemes are derived that preserve the  positivity and the trace of  the density operator, i.e. of the quantum state.
\end{abstract}

 Positivity and trace preserving numerical scheme.

\section{Intoduction}

\begin{figure}[h]
  \centerline{ \includegraphics[width=\textwidth]{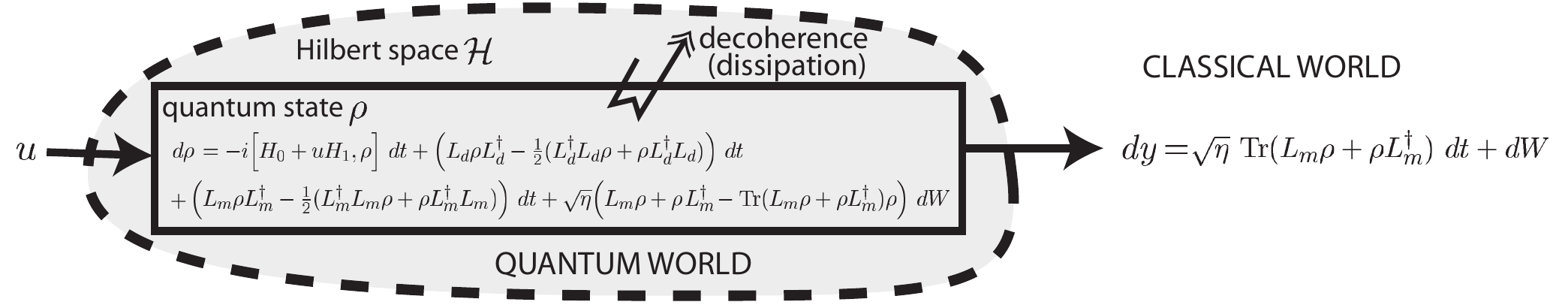}}
\caption{Classical   Kalman state-space descriptions are  replaced, for quantum systems,  by   Stochastic Master Equation (SME)  descriptions where the measurement back-action is revealed  here by the same Wiener process $W$ shared by the quantum state $\rho$ and by  the output $y$. }  \label{fig:ClassInOut}
\end{figure}

An increasing number of experiments controlling   quantum states are  conducted  with  various physical supports such as spins,  atoms, trapped ions,  photons,   superconducting circuits, electro-mechanical circuits,  optomechanical cavities~(see, e.g., \cite{haroche-raimondBook06,QuantumMachinesHouches2011,AspelmeyerKippenbergMarquardt2014,GardinerZollerBookTwo2015,BovwenMilburnBook2016}).  As illustrated    on Fig.~\ref{fig:ClassInOut}, {  the quantum dynamics of these experiments  can be precisely  described by well structured stochastic differential equations,  called   Stochastic Master Equations (SME)}. They  govern the  relationships  between the   input  $u$ corresponding to the classical parameters manipulated by the experimentalists and the classical output  $y$ corresponding to the  observed measurements. These  SME are expressed with  operators  for which  non-commutative calculus and   commutation relationships play a fundamental role. These   SME   are   the quantum analogue of the  classical  Kalman state-space descriptions,  $\dotex x = A x+Bu+ w$ and  $y= C x+ v $,   with noise $(w,v)$~\cite{kailath-book}.

For Fig.~\ref{fig:ClassInOut}, with  classical input $u$ and classical output $y$, this SME  reads (diffusive case  with It\={o} formulation~\cite{BarchielliGregorattiBook}, $L_\nu^{\dag}$ stands for Hermitian conjugate of operator $L_\nu$):
\begin{multline} \label{eq:SME}
d\rho_t  = - i \Big[H_0+uH_1,\rho_t\Big]~ dt  + \left(\sum_{\nu=d,m}L_\nu \rho_t L_\nu^\dag - \tfrac{1}{2}( L_\nu^\dag L_\nu \rho_t + \rho_t  L_\nu^\dag L_\nu)\right)~dt +\ldots
\\
\ldots  + \sqrt{\eta} \Big( L_m \rho_t + \rho_t L_m^\dag  - \text{Tr}(L_m\rho_t+\rho_t L_m^\dag ) \rho_t \Big) ~ dW_t
 ,
\end{multline}
where the same Wiener process $W_t$  is shared by the  state dynamics  and the output map
\begin{equation} \label{eq:dy}
  dy_t=\sqrt{\eta}~\text{Tr}(L_m\rho_t+\rho_t L_m^\dag) ~dt + dW_t
  .
\end{equation}
The state $\rho$ is a density operator (a self-adjoint,  non-negative  and trace-one  operator)  on a Hilbert space $\mathcal{H}$. Its dynamics~\eqref{eq:SME} are parameterized here via two  self-adjoint  operators (Hamiltonians) $H_0$ and $H_1$ ($[\cdot,\cdot]$ stands for the commutator)  and two Lindblad operators, $L_d$  describing a  decoherence channel and  $L_m$  a   measurement channel   of  efficiency $\eta\in[0,1]$.  When    $\eta=0$, $\rho$ follows a deterministic linear  master equation, called {Lindblad master equation}  with two decoherence channels described by $L_d$ and $L_m$,
\begin{equation}\label{eq:Lindblad}
\dotex \rho = - i \Big[H_0+uH_1,\rho\Big] + \sum_{\nu=d,m}L_\nu \rho L_\nu^\dag - \tfrac{1}{2}( L_\nu^\dag L_\nu \rho + \rho  L_\nu^\dag L_\nu)
\end{equation}
and the measurement $y_t=W_t$ boils down to a  Wiener process without any relations with $u$ and $\rho$ and thus can be discarded. Notice that~\eqref{eq:Lindblad} corresponds also to the ensemble average dynamics of the SME~\eqref{eq:SME}. Notice also that the initial value problems (Cauchy problems)  attached to~\eqref{eq:SME} or to~\eqref{eq:Lindblad} are non trivial  mathematical problems when the dimension of the underlying Hilbert space $\mathcal{H}$ is infinite and  the Hamiltonian or Lindblad  operators are unbounded  (see e.g \cite{AlickiLendiBook2007,TarasovBook2008}).


{These SME  rely  on the well developed theory of open quantum systems combining  irreversibility  due to decoherence (quantum dissipation)~\cite{daviesBook1976,BreuerPetruccioneBook} and stochasticity due to measurement  back-action~\cite{wiseman-milburnBook,CarmichaelBook1993,DalibCM1992PRL,JacobsSteck2006,JacobsBook2014}.}   More general  SME  than the one depicted on Fig.~1,   with several Lindblad  operators and/or  driven  by Poisson processes  (counting  measurement),  admit similar structures~\cite{RouchonSeoul2014}. Even if the initial system is known to be non Markovian, it is always possible in general to adjunct a dynamical model of the environment and  to recover a Markovian model with an  SME structure but  of larger dimension \cite[part IV]{BreuerPetruccioneBook}.   For composite systems made of several interacting  quantum sub-systems  such SME  models   are also  derived from a  quantum network  theory~\cite{GoughJ2009ACITo,GoughJ2009} gathering in a concise way,  quantum stochastic calculus~\cite{ParthasarathyBook92}, Heisenberg description of   input/output theory~\cite{YurkeD1984PRA,gardiner-zollerBook} and quantum filtering~\cite{Belavkin1992}.

The goal of this paper  is to provide an introduction to the structure of  these quantum SME illustrated via  a composite system made of two key quantum sub-systems (qubits and photons, see~\ref{ap:qubitphoton}) and  based on  three fundamental  quantum  rules  (unitary evolution derived from Schr\"{o}dinger equation, measurement back-action with the collapse of the wave-packet, composite systems relying on tensor products, see~\ref{ap:3Qrules}).

\begin{figure}
  \centerline{\includegraphics[width=0.8\textwidth]{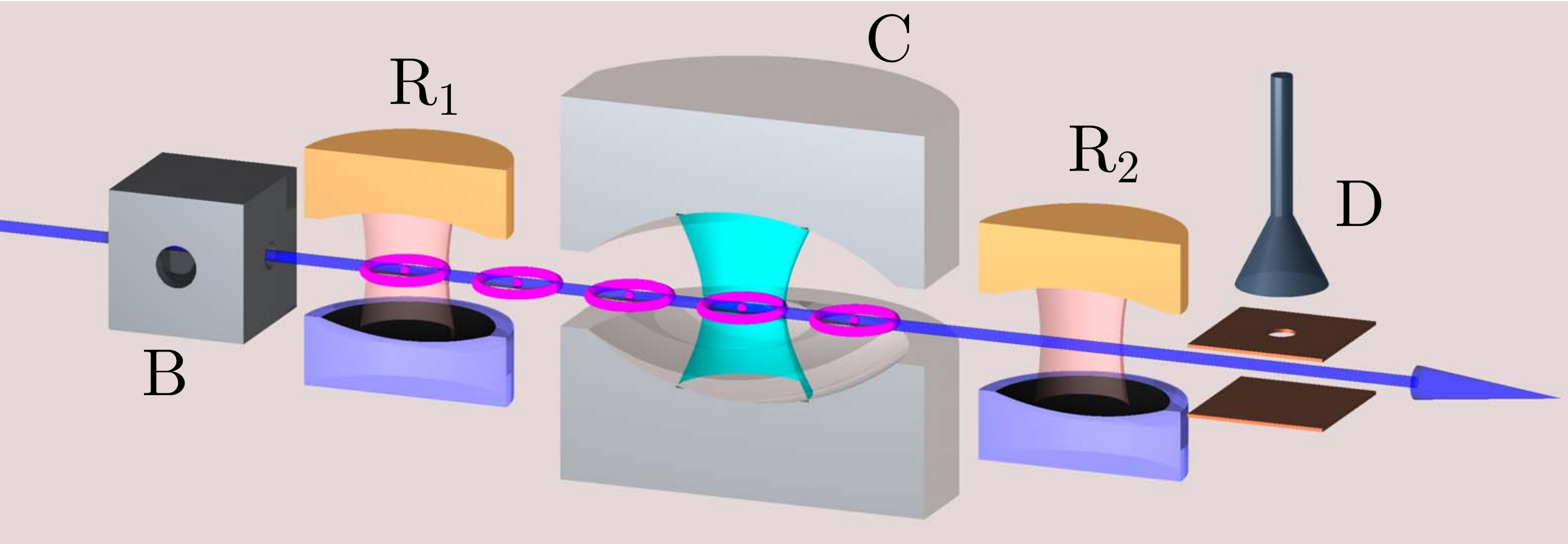}}
  \caption{Scheme of the LBK experiment where photons   are observed via probe atoms.
   The photons in blue  are trapped between the two mirrors of the cavity $C$. They are  probed by  two-level  atoms (the small pink torus) flying out the preparation box $B$, passing  through the cavity  $C$ and measured in $D$. Each atom is manipulated  before and after $C$ in  Ramsey cavities $R_1$ and  $R_2$, respectively.   It is  finally  detected in $D$  either in ground state $\ket g$   or in excited state   $\ket e$.    }\label{fig:LKBsetup0}
\end{figure}

Section~\ref{sec:discrete} is devoted to discrete-time formulation of SME. One starts with  the Markov chain modelling  the LKB\footnote{LKB for Laboratoire Kastler Brossel.} photon box of Fig.~\ref{fig:LKBsetup0} \cite{haroche-raimondBook06}  where photons are measured by probe atoms described by two-level systems, i.e. qubits.  Two kinds of interactions between the photons and the atoms are considered (see~\ref{ap:qubitphoton} for  operator notations) :
\begin{itemize}
  \item dispersive interaction leading to Quantum Non Demolition (QND) measurement of photons; the qubit/photon  interaction  is dispersive where
$
\bH_{int}= -\chi \big( \ketbra{e}{e} - \ketbra{g}{g}\big) \otimes \bn
$  (with $\chi$ a constant parameter)
 yields  $\bU_{\theta=\chi  T}= e^{-i  T\bH_{int}}$,  the Schr\"{o}dinger propagator during the time  $T>0$,  given by the explicit formula:
 \begin{equation}\label{eq:dispersivePropagator}
   \bU_{\theta} =\ketbra{g}{g}\otimes e^{-i \theta\bn}+ \ketbra{e}{e}\otimes  e^{i \theta\bn}, \quad \theta= \chi T
 \end{equation}
 where $\theta=\chi T$.

  \item resonant interaction   stabilizing then the photons in vacuum state;
   the qubit/photon  interaction is here resonant  where
$
\bH_{int}=  i \frac{\omega}{2}\Big(\ketbra{g}{e} \otimes \ba^\dag - \ketbra{e}{g}\otimes \ba\Big)
$  (with $\omega$ a constant parameter)
 yields  $\bU_{\theta=\omega  T}= e^{-i  T\bH_{int}}$,  the Schr\"{o}dinger propagator during the time  $T>0$,  given by the explicit formula:
 \begin{multline}\label{eq:resonantPropagator}
   \bU_{\theta} = \ketbra{g}{g}\otimes \cos( \theta \sqrt{\bn})+ \ketbra{e}{e}\otimes \cos( \theta \sqrt{\bn+\bI})
\\+\ketbra{g}{e} \otimes\frac{\sin( \theta \sqrt{\bn})}{ \sqrt{\bn}} \ba^\dag
-\ketbra{e}{g}\otimes \ba  \frac{\sin( \theta \sqrt{\bn})}{ \sqrt{\bn}}
 \end{multline}
 where $\theta=\omega T$.
\end{itemize}
One explains on this key system  how to take into account  measurement errors   and why  the  density operator as quantum state is then  crucial.  One concludes section~\ref{sec:discrete} with the  general structure  of discrete-time SME governing the stochastic dynamics of  open quantum systems subject to  unperfect measurement and decoherence.

Section~\ref{sec:continuous} is devoted to continuous-time SME. One considers here the reverse situation where the qubit is measured by probe photons. Dispersive interaction, measurement of one quadrature $\bQ$ of the photons (an observable with a continuous spectrum)  and $\theta=\chi T$ scaling as $\sqrt{dt}$ yields  a continuous-time SME driven by a Wiener process  of form~\eqref{eq:SME} with $L_d=0$, $L_m\propto \Sz$ and  $\eta=1$.  One shows how  measurement errors  tend to decrease $\eta$ towards  $0$. Then the general structure of diffusive SME is  presented with an equivalent Kraus-map formulation yielding  alinear time-integration numerical scheme preserving the positivity and the trace.
Resonant interaction, measurement of the photon-number operator  $\bn$  and $\theta=\omega T$ scaling as $dt$ yield  a continuous-time SME driven by a Poisson  process associated to  the  measurement counter. One shows how to include  measurement  imperfections and gives the  structure of continuous-time SME  driven by Poisson processes. Last subsection of section~\ref{sec:continuous} provides a very general structure of continuous-time SME driven by Wiener and  Poisson processes governing the stochastic dynamics of   open quantum systems subject to diffusive/counting  unperfect measurements and decoherence.

The conclusion section~\ref{sec:conclusion} provides some   comments  and  references  related to feedback, parameter estimation and filtering issues. These comments and references  are far from being  exhaustive.

\section{Discrete-time  formulation} \label{sec:discrete}

\subsection{Photons measured by qubits (dispersive interaction)}

The wave function of the photon is denoted here by  $\ket{\psi}$. From the scheme of Fig.~\ref{fig:LKBsetup0}, the qubit produced by the $B$ in state $\ket{g}$   is subject in $R_1$ to a rotation of $\pi/4$ in the plane $\text{span}\{\ket{g},\ket{e}\}$, then  interacts during $T$  dispersively with the photons in $C$, is subject to  the reverse  rotation of  $-\pi/4$ in $R_2$ and finally is measured in $D$ according to $\Sz$.  The Schr\"{o}dinger evolution of the  qubit/photon wave function $\ket{\Psi}$ between $B$ and just before $D$ is given by the following unitary evolution $\bU$:
\begin{multline*}
\bU=
 \left(  \left( \left(\tfrac{\ket{g} -\ket{e}}{\sqrt{2}}\right) \bra{g} +  \left(\tfrac{\ket{g} +\ket{e}}{\sqrt{2}}\right) \bra{e} \right) \otimes \bI \right) \\
\Big(\ketbra{g}{g} \otimes e^{-i \theta\bn}+ \ketbra{e}{e}\otimes   e^{i \theta\bn} \Big) \\ \left( \left( \left(\tfrac{\ket{g} +\ket{e}}{\sqrt{2}}\right) \bra{g} +  \left(\tfrac{-\ket{g} +\ket{e}}{\sqrt{2}}\right) \bra{e} \right)
\otimes \bI \right)
.
\end{multline*}
Applied  to the  value of $\ket{\Psi} = \ket{g} \otimes \ket{\psi}$ when the qubit   leaves $B$, one gets\footnote{Tensor sign $\otimes$ and tensor  product with identity operator $\bI$ are   not explicitly recalled in the formula  as  it is usually done  when there is no ambiguity:
$\ket{g} \otimes \ket{\psi}$ and
$\ketbra{g}{g} \otimes e^{-i \theta\bn}+ \ketbra{e}{e} \otimes  e^{i \theta\bn}$ read then $\ket{g}  \ket{\psi}$ and  $\ketbra{g}{g} e^{-i \theta\bn}+ \ketbra{e}{e}  e^{i \theta\bn} $; similarly $\left( \left(\tfrac{\ket{g} -\ket{e}}{\sqrt{2}}\right) \bra{g} +  \left(\tfrac{\ket{g} +\ket{e}}{\sqrt{2}}\right) \bra{e} \right) \otimes \bI$ just becomes $\left( \left(\tfrac{\ket{g} -\ket{e}}{\sqrt{2}}\right) \bra{g} +  \left(\tfrac{\ket{g} +\ket{e}}{\sqrt{2}}\right) \bra{e} \right)$.
}
$$
 \bU ~(\ket{g}  \ket{\psi})
= \ket{g} ~\cos(\theta\bn) \ket{\psi}+   \ket{e}~ i\sin(\theta\bn) \ket{\psi}
.
$$

Measuring in $D$, the observable $\Sz$ yields   the collapse of $\ket{\Psi}$ into a separable state  $ \ket{g} ~\cos(\theta\bn) \ket{\psi}$  or  $ \ket{e} ~i\sin(\theta\bn) \ket{\psi}$, eigen-vectors of $\Sz\otimes \bI \equiv \Sz$  with eigenvalues $-1$ or  $1$, respectively.  Numbering the qubit by the integer  $k$ and removing the  qubit state, one gets the following Markov process induced by the passage of  qubit number $k$:
$$
\ket{\psi_{k+1}} = \left\{
                     \begin{array}{ll}
                      \frac{\cos(\theta\bn) \ket{\psi_k}}{\sqrt{\bra{\psi_k}\cos^2(\theta\bn)\ket{\psi_k}}} \, & \hbox{if $y_k=g$ with probability $\bra{\psi_k}\cos^2(\theta\bn)\ket{\psi_k}$ ;} \\
  \frac{i\sin(\theta\bn) \ket{\psi_k}}{\sqrt{\bra{\psi_k}\sin^2(\theta\bn)\ket{\psi_k}}} \, & \hbox{if $y_k=e$ with probability $\bra{\psi_k}\sin^2(\theta\bn)\ket{\psi_k}$ ;} \\
                     \end{array}
                   \right.
$$
where $y_k\in\{g,e\}$ is the classical signal produced by the quantum measurement of  qubit $k$.
The density operator  formulation of this Markov process  reads  ($\rho \equiv \ket{\psi}\bra{\psi}$ ):
$$
\rho_{k+1} = \left\{
                     \begin{array}{ll}
                      \frac{\bM_g \rho_k \bM_g^\dag}{\Tr{\bM_g \rho_k \bM_g^\dag}} \, & \hbox{if $y_k=g$ with probability $\Tr{\bM_g \rho_k \bM_g^\dag}$ ;} \\
                      \frac{\bM_e \rho_k \bM_e^\dag}{\Tr{\bM_e \rho_k \bM_e^\dag}} \, & \hbox{if $y_k=e$ with probability $\Tr{\bM_e \rho_k \bM_e^\dag}$ ;}
                     \end{array}
                   \right.
$$
with measurement Kraus operators $\bM_g =\cos(\theta\bn)$ and $\bM_e=\sin(\theta\bn)$. Notice that $\bM_g^\dag \bM_g + \bM_e^\dag \bM_e = \bI$.

When $\theta/\pi$ is irrational, each realization of this  Markov process, starting from $\rho_0$ satisfying  $\rho_0 \ket{n} =0$ for $n$ large enough, converges almost surely  towards a Fock state $\ket{\bar n}\bra{\bar n}$ for some $\bar n$. More precisely, the probability that a realisation  converges towards $\ket{\bar n}\bra{\bar n}$ is given by the initial population $\bra{\bar n}\rho_0\ket{\bar n}$ (see, e.g.,\cite{HarocheBruneRaimondJPh92,Bauer2011,AminiSDSMR2013A}).
This almost sure  convergence  can be seen from the following Lyapunov function (super-martingale)
$$
V(\rho)= \sum_{0\leq n_1 < n_2} \sqrt{\bra{n_1}\rho\ket{n_1} \bra{n_2}\rho\ket{n_2}}
$$
that converges in average  towards $0$  since its expectation value from  step to step satisfies :
$$
\EE{V(\rho_{k+1})}{\rho_k} \leq \left(\max_{0\leq n_1 < n_2} |\cos(\theta(n_1\pm n_2)|\right) ~ V(\rho_k)
.
$$

\subsection{Photons measured by qubits (resonant interaction)}

The photon wave function is still denoted by  $\ket{\psi}$. The  qubit  coming from box $B$ of Fig.~\ref{fig:LKBsetup0} is  in  $\ket{g}$. The Ramsey  zones $R_1$ and $R_2$ are inactive. The  resonant interaction during the passage of the qubit in $C$  yields    the propagator~\eqref{eq:resonantPropagator}. Thus, the wave function $\ket{\Psi}$ of the composite qubit/photon system,   just before the qubit measurement  in $D$, is as follows:
\begin{multline*}
\Bigg(\ketbra{g}{g} \cos( \theta \sqrt{\bn})+ \ketbra{e}{e} \cos( \theta \sqrt{\bn+\bI})
\\
+\ketbra{g}{e} \frac{\sin( \theta \sqrt{\bn})}{ \sqrt{\bn}} \ba^\dag
-\ketbra{e}{g} \ba  \frac{\sin( \theta \sqrt{\bn})}{ \sqrt{\bn}}\Bigg) \ket{g}\ket\psi
\\
= \ket{g} ~ \cos( \theta \sqrt{\bn}) \ket{\psi}-   \ket{e}~\ba  \frac{\sin( \theta \sqrt{\bn})}{ \sqrt{\bn}} \ket{\psi}
\end{multline*}

Therefore, the resulting Markov process  associated to the measurement of the observable   $\Sz= \ketbra{e}{e} - \ketbra{g}{g}$  with classical signal  $y\in\{g,e\}$ is as follows:
$$
\ket{\psi_{k+1}} = \left\{
                     \begin{array}{ll}
                      \frac{\cos( \theta \sqrt{\bn}) \ket{\psi_k}}{\sqrt{\bra{\psi_k}\cos^2( \theta \sqrt{\bn})\ket{\psi_k}}} \,
& \hbox{if $y_k=g$ with probability $\bra{\psi_k}\cos^2( \theta \sqrt{\bn})\ket{\psi_k}$ ;} \\
 - \frac{\ba  \frac{\sin( \theta \sqrt{\bn})}{ \sqrt{\bn}} \ket{\psi_k}}{\sqrt{\bra{\psi_k}\sin^2( \theta \sqrt{\bn})\ket{\psi_k}}}
 \, & \hbox{if $y_k=e$ with probability $\bra{\psi_k}\sin^2( \theta \sqrt{\bn})\ket{\psi_k}$ ;} \\
                     \end{array}
                   \right.
$$
The corresponding density operator formulation is then
$$
\rho_{k+1} = \left\{
                     \begin{array}{ll}
                      \frac{\bM_g \rho_k \bM_g^\dag}{\Tr{\bM_g \rho_k \bM_g^\dag}} \, & \hbox{if $y_k=g$ with probability $\Tr{\bM_g \rho_k \bM_g^\dag}$ ;} \\
                      \frac{\bM_e \rho_k \bM_e^\dag}{\Tr{\bM_e \rho_k \bM_e^\dag}} \, & \hbox{if $y_k=e$ with probability $\Tr{\bM_e \rho_k \bM_e^\dag}$ ;}
                     \end{array}
                   \right.
$$
with measurement Kraus operators $\bM_g =\cos( \theta \sqrt{\bn})$ and $\bM_e=\ba \frac{\sin( \theta \sqrt{\bn})}{ \sqrt{\bn}}$. Notice that, once again, $\bM_g^\dag \bM_g + \bM_e^\dag \bM_e = \bI$.

When $\theta\sqrt{n}/\pi$ is irrational for all  positive integer $n$, this Markov process converges almost surely towards   vacuum  state  $\ket{0}\bra{0}$   when $\rho_0\ket{n}=0$ for $n$ large enough. This results from the following the Lyapunov function (super-martingale)
$$
V(\rho)=\Tr{\bn \rho}
$$ since
$$
\EE{V(\rho_{k+1})}{\rho_k} = V(\rho_{k}) - \Tr{\sin^2( \theta \sqrt{\bn}) \rho_k}
.
$$

\subsection{Measurement errors}
In presence of measurement imperfections and errors, one has to update the quantum $\rho$ according to Bayes rule by taking  as quantum state, the expectation value of $\rho_{k+1}$ given by
$$
\rho_{k+1} = \left\{
                     \begin{array}{ll}
                      \frac{\bM_g \rho_k \bM_g^\dag}{\Tr{\bM_g \rho_k \bM_g^\dag}} \, & \hbox{if $y_k=g$ with probability $\Tr{\bM_g \rho_k \bM_g^\dag}$ ;} \\
                      \frac{\bM_e \rho_k \bM_e^\dag}{\Tr{\bM_e \rho_k \bM_e^\dag}} \, & \hbox{if $y_k=e$ with probability $\Tr{\bM_e \rho_k \bM_e^\dag}$ ;}
                     \end{array}
                   \right.
$$
 knowing $\rho_k$ and the information provides by the imperfect measurement outcome.
 Assume firstly, that the detector $D$ is broken. Then, we get the following  linear, trace-preserving and completely positive map
$$
\rho_{k+1} = \KK(\rho_k) \triangleq \EE{\rho_{k+1}}{\rho_k}=  \bM_g \rho_k \bM_g^\dag + \bM_e \rho_k \bM_e^\dag
$$
called in quantum information a  quantum channel (see~\cite{nielsen-chang-book}).

 When  the qubit detector $D$,  producing the classical measurement signal $y_k\in\{g,e\}$,  has symmetric errors characterized by the single error rate  $\eta \in (0,1)$, the probability of detector  outcome  $g$ (resp. $e$) knowing that the perfect outcome  is $e$ (resp. $g$), Bayes law gives directly
{\small
\begin{equation}\label{eq:discreteSME}
\hspace{-2em}\rho_{k+1} =   \left\{
                     \begin{array}{lrlr}
                     \EE{\rho_{k+1}}{y_k=g,\rho_k} = \frac{ (1-\eta)\bM_g \rho_k \bM_g^\dag +\eta \bM_e \rho_k \bM_e^\dag  }{\Tr{(1-\eta)\bM_g \rho_k \bM_g^\dag +\eta \bM_e \rho_k \bM_e^\dag }} \\ \qquad\text{with probability } \PP(y_k=g|\rho_k)=\Tr{(1-\eta)\bM_g \rho_k \bM_g^\dag +\eta \bM_e \rho_k \bM_e^\dag },  \\[1.em]
                      \EE{\rho_{k+1}}{y_k=e,\rho_k} = \frac{ \eta\bM_g \rho_k \bM_g^\dag +(1-\eta) \bM_e \rho_k \bM_e^\dag  }{\Tr{\eta\bM_g \rho_k \bM_g^\dag +(1-\eta) \bM_e \rho_k \bM_e^\dag }}
                      \\ \qquad\text{with probability }  \PP(y_k=e|\rho_k)=\Tr{\eta\bM_g \rho_k \bM_g^\dag +(1-\eta) \bM_e \rho_k \bM_e^\dag }
                     \end{array}
                   \right.
\end{equation}
}
Notice that a broken detector corresponds to  $\eta=1/2$ and one recovers the above quantum channel.

\subsection{Stochastic Master Equation (SME) in discrete-time}

In fact, the general structure of discrete-time SME  can always  be constructed  from the knowledge of a   quantum channel (trace preserving completely positive map) having the following  Kraus decomposition (which is not unique)
$$
\KK(\rho) =\sum_{\mu} \bM_\mu \rho \bM_\mu^\dag \quad\text{where  $\sum_{\mu} \bM_\mu^\dag \bM_\mu= \bI$}
$$
and a left stochastic matrix $(\eta_{y,\mu})$ where $y$ corresponds to the different imperfect measurement outcomes.
Set  $\KK_{y}(\rho)=\sum_{\mu} \eta_{y,\mu}  \bM_\mu \rho \bM_\mu^\dag$.
The SME associated to $\KK$ and $\eta$ reads
$$
\rho_{k+1} = \frac{\KK_{y_k}(\rho_k) }{ \Tr{\KK_{y_k}(\rho_k)}} \quad\text{where $y_k=y$ with probability $\Tr{\KK_y(\rho_k)}$}
$$
Notice that $\KK= \sum_y \KK_y$ since $\eta$ is left stochastic.
Here the Hilbert space $\mathcal{H}$  is arbitrary and can be of infinite dimension, the Kraus operator $\bM_\mu$ are bounded operator on $\mathcal{H}$  and $\rho$ is a density operator on $\mathcal{H}$ (Hermitian, trace-class with trace one,  non-negative).

To recover the previous discrete-time  SME~\eqref{eq:discreteSME},  use the above general formulae with the  quantum channel  $\KK(\rho)= \bM_g\rho \bM_g^\dag + \bM_e\rho \bM_e^\dag$ and the  left stochastic matrix
$$
\left(
       \begin{array}{cc}
         \eta_{g,g}=1-\eta_e  & \eta_{g,e}=\eta_g \\
         \eta_{e,g}= \eta_e & \eta_{e,e}=1-\eta_g \\
       \end{array}
     \right)
$$
where $\eta_g$ (resp. $\eta_e$)  is the error probability  associated to outcome $g$ (resp. $e$). Notice that in~\eqref{eq:discreteSME} the error model is symmetric with $\eta_g=\eta_e$ corresponding to $\eta\in[0,1]$.

\section{Continuous-time formulation} \label{sec:continuous}

Contrarily to~section~\ref{sec:discrete},   photons measure here a qubit.

\subsection{Qubits measured by photons: dispersive interaction and discrete-time}

The qubit wave function is denoted by $\ket{\psi}$. The photons,  before interacting with  the qubit,  are in the  coherent state $\ket{i \tfrac{\alpha}{\sqrt{2}}}$ with $ \alpha$ real and strictly positive.  The interaction is dispersive according to~\eqref{eq:dispersivePropagator}.
After the interaction and just before the measurement performed on the photons, the composite qubit/photon wave function $\ket{\Psi}$ reads:
$$
\Big( \ketbra{g}{g} e^{-i \theta\bn}+ \ketbra{e}{e}  e^{i \theta\bn}\Big) \ket\psi \ket{i \tfrac{\alpha}{\sqrt{2}}}
= \braket{g}{\psi} \ket{g} ~ \ket{i e^{-i\theta} \tfrac{\alpha}{\sqrt{2}}} + \braket{e}{\psi} \ket{e} ~ \ket{i e^{i\theta}\tfrac{\alpha}{\sqrt{2}}}
$$
since for any coherent state $\ket{\beta}$ of complex amplitude $\beta$,   $e^{i\theta \bn} \ket{\beta}$ is also a coherent state of complex  amplitude
$e^{i\theta} \beta$ ($e^{i\theta \bn} \ket{\beta}=\ket{e^{i\theta}\beta}$).

Assume that the perfect measurement outcome  $y$ belongs to $\mathbb{R}$  and  corresponds to the phase-plane  observable $\bQ=\frac{\ba+\ba^\dag}{\sqrt{2}}$ having  the entire real line as spectrum. Its spectral decomposition reads formally  $\bQ= \int_{-\infty}^{+\infty} q \ketbra{q}{q} dq $ where $\ket{q}$ is  the wave function associated to the  eigen-value $q$. $\ket{q}$  is not  a usual wave function, i.e., in $L^2(\RR,\CC)$, but one has formally $\braket{q}{q'}=\delta(q-q')$  (see, e.g., \cite{barnett-radmoreBook}).
Since
$$
\ket{i e^{\pm i\theta}\tfrac{\alpha}{\sqrt{2}}}=\tfrac{1}{\pi^{1/4}} \int_{-\infty}^{+\infty}  e^{iq \alpha\cos\theta} e^{-\frac{(q\pm  \alpha \sin\theta)^2}{2}} \ket{q} dq,
$$
we have
\begin{multline*}
  \braket{g}{\psi} \ket{g} ~ \ket{i e^{-i\theta}\tfrac{\alpha}{\sqrt{2}}} + \braket{e}{\psi} \ket{e} ~ \ket{i e^{i\theta}\tfrac{\alpha}{\sqrt{2}}}
\\
=\tfrac{1}{\pi^{1/4}}  \int_{-\infty}^{+\infty} e^{iq \alpha\cos\theta} \left( e^{-\frac{(q- \alpha \sin\theta)^2}{2}} \braket{g}{\psi} \ket{g} + e^{-\frac{(q+ \alpha \sin\theta)^2}{2}} \braket{e}{\psi} \ket{e} \right)   \ket{q} dq
.
\end{multline*}
Thus
$$
\ket{\psi_{k+1}} = e^{iy_k \alpha\cos\theta}\frac{e^{-\frac{(y_k- \alpha \sin\theta)^2}{2}} \braket{g}{\psi_k} \ket{g} + e^{-\frac{(y_k+ \alpha \sin\theta)^2}{2}} \braket{e}{\psi_k} \ket{e}}{ \sqrt{e^{-(y_k- \alpha \sin\theta)^2} |\braket{g}{\psi_k}|^2 +  e^{-(y_k+\alpha \sin\theta)^2} |\braket{e}{\psi_k}|^2}}
$$
where  $y_k\in[y,y+dy]$ with probability $\frac{e^{-(y- \alpha \sin\theta)^2} |\braket{g}{\psi_k}|^2 +  e^{-(y+\alpha \sin\theta)^2} |\braket{e}{\psi_k}|^2}{\sqrt{\pi}} dy$.

The density operator formulation reads then
$$
\rho_{k+1} =  \frac{\bM_{y_k} \rho_k \bM_{y_k}^\dag}{\Tr{\bM_{y_k} \rho_k \bM_{y_k}^\dag}}
\quad\text{where  $y_k\in[y,y+dy]$ with probability $\Tr{\bM_{y} \rho_k \bM_{y}^\dag}dy$}
$$
and  measurement Kraus operators
$$
\bM_y = \tfrac{1}{\pi^{1/4}} e^{-\frac{(y- \alpha \sin\theta)^2}{2}} \ketbra{g}{g}+\tfrac{1}{\pi^{1/4}} e^{-\frac{(y+ \alpha \sin\theta)^2}{2}} \ketbra{e}{e}.
$$
Notice that
\begin{equation}\label{eq:TRM}
\Tr{\bM_y \rho \bM_y^\dag}= \frac{1}{\sqrt{\pi}} e^{-(y- \alpha \sin\theta)^2}
\bra{g}\rho\ket{g} + \frac{1}{\sqrt{\pi}}  e^{-(y+ \alpha \sin\theta)} \bra{e}\rho\ket{e}
\end{equation}
 and $\int_{-\infty}^{+\infty} \bM_y^\dag \bM_y ~dy = \ketbra{g}{g}+\ketbra{e}{e}= \bI$.

For  $\alpha \neq 0$, one has almost sure convergence  towards $\ket{g}$ or $\ket{e}$ deduced from the following    Lyapunov function
$$
V(\rho)= \sqrt{\bra{g}\rho\ket{g} \bra{e}\rho\ket{e}}
$$
and
$$
\EE{V(\rho_{k+1})}{\rho_k}= e^{-\alpha^2\sin^2\theta} ~ V(\rho_k)
.
$$

Assume that the detection  of $y$ is not perfect. The probability density of  $y$ knowing that the perfect  detection is $q$ is also a Gaussian given by $\frac{1}{\sqrt{\pi\sigma}} e^{-\frac{(y-q)^2}{\sigma}}$ for some error  parameter $\sigma >0$.
Then the above  Markov process  becomes
$$
\rho_{k+1} =  \frac{\KK_{y_k}(\rho_k)  }{ \Tr{\KK_{y_k}(\rho_k)}}
$$
where
$$
\KK_{y}(\rho)= \int_{-\infty}^{\infty} \tfrac{1}{\sqrt{\pi\sigma}} e^{-\frac{(y-q)^2}{\sigma}} \bM_q \rho \bM_q^{\dag}~ dq
$$
Standard computations show that
\begin{multline*}
 \KK_{y}(\rho) = \tfrac{1}{\sqrt{\pi(1+\sigma)}}
\left(
 e^{-\frac{(y-\alpha\sin\theta)^2}{1+\sigma}}  \bket{g|\rho|g}\ketbra{g}{g} +
 e^{-\frac{(y+\alpha\sin\theta)^2}{1+\sigma}}  \bket{e|\rho|e}\ketbra{e}{e} \right.
\\ \left. +
 e^{-\frac{y^2}{1+\sigma} - (\alpha\sin\theta)^2 }  \big( \bket{e|\rho|g}\ketbra{e}{g} +\bket{g|\rho|e}\ketbra{g}{e}\big)
\right)
.
\end{multline*}

\subsection{Dispersive interaction and continuous-time diffusive limit}

Consider the above Markov process with perfect detection $y\in\RR$. From~\eqref{eq:TRM}, one gets
$$
\EE{y_k}{\rho_k=\rho}\triangleq \overline{y}= -\alpha \sin\theta ~\Tr{\Sz \rho}, ~
\EE{y_k^2}{\rho_k=\rho}\triangleq\overline{y^2}=1/2+(\alpha\sin\theta)^2
.
$$
When $0<\alpha\sin\theta = \epsilon \ll 1$, we have up-to third order terms versus $\epsilon y$,
\begin{multline*}
\frac{\bM_y \rho \bM_y^\dag}{\Tr{\bM_y \rho \bM_y^\dag}}=
 \frac{(\cosh(\epsilon y) - \sinh(\epsilon y) \Sz) \rho (\cosh(\epsilon y) - \sinh(\epsilon y) \Sz)}{
 \cosh(2\epsilon y) - \sinh(2\epsilon y) \Tr{\Sz\rho}}
\\
\approx \frac{\rho - \epsilon y (\Sz \rho + \rho \Sz) +(\epsilon y)^2( \rho + \Sz\rho \Sz)}{1- 2 \epsilon y \Tr{\Sz \rho} +2 (\epsilon y)^2 }
\\
= \rho + (\epsilon y)^2 \Big(\Sz \rho \Sz -\rho\Big) +\Big(\Sz \rho + \rho \Sz - 2\Tr{\Sz\rho} \rho\Big) \Big(-\epsilon y -2 (\epsilon y)^2 \Tr{\Sz \rho}\Big)
.
\end{multline*}
Replacing $\epsilon^2 y^2$ by its expectation value   independent of $\rho$ one gets, up to third order terms versus $\epsilon y$ and $\epsilon$:
$$
\frac{\bM_y \rho \bM_y^\dag}{\Tr{\bM_y \rho \bM_y^\dag}}
\approx \rho + \tfrac{\epsilon^2}{2} \Big(\Sz \rho \Sz -\rho\Big) +\Big(\Sz \rho + \rho \Sz - 2\Tr{\Sz\rho} \rho\Big) \Big(-\epsilon y -\epsilon^2 \Tr{\Sz \rho}\Big)
.
$$
Set $\epsilon^2=2dt$ and $\epsilon y = -2 \Tr{\Sz\rho}dt -  dW$.
Since  by construction
$$
\EE{\epsilon y_k}{\rho_k=\rho}= -\epsilon^2 \Tr{\Sz \rho} \text{ and } \EE{(\epsilon y_k)^2}{\rho_k=\rho}= \epsilon^2 + \epsilon^4
$$   one has $\EE{dW}{\rho} = 0$ and $\EE{dW^2}{\rho}=dt$ up to order $4$ versus $\epsilon$.
 Thus for $dt$ very small, we recover the following diffusive  SME
$$
\rho_{t+dt} = \rho_t + dt  \Big(\Sz \rho_t \Sz -\rho\Big)+\Big(\Sz \rho_t + \rho_t \Sz - 2\Tr{\Sz\rho_t} \rho\Big) \Big(dy_t  - 2   \Tr{\Sz \rho_t} dt\Big)
$$
with $dy_t = 2\Tr{\Sz \rho_t} dt + dW_t$ replacing  $-\epsilon y$  and $dy_t^2=dW_t^2=dt$  according to Ito rules.

With  measurement errors parameterized by $\sigma >0$, the partial Kraus map
\begin{multline*}
 \KK_{y}(\rho) = \tfrac{1}{\sqrt{\pi(1+\sigma)}}
\left(
 e^{-\frac{(y-\epsilon)^2}{1+\sigma}}  \bket{g|\rho|g}\ketbra{g}{g} +
 e^{-\frac{(y+\epsilon)^2}{1+\sigma}}  \bket{e|\rho|e}\ketbra{e}{e} \right.
\\ \left. +
 e^{-\frac{y^2}{1+\sigma} -\epsilon^2 }  \big( \bket{e|\rho|g}\ketbra{e}{g} +\bket{g|\rho|e}\ketbra{g}{e}\big)
\right)
\end{multline*}
yields
$$
\EE{y_k}{\rho_k}\triangleq \overline{y}=-\epsilon\Tr{\Sz\rho}, ~
\EE{y_k^2}{\rho_k}\triangleq\overline{y^2}=(1+\sigma)/2+\epsilon^2
.
$$
From
\begin{multline*}
\KK_y(\rho)
 \\
= \frac{e^{-\frac{y^2+\epsilon^2}{1+\sigma} }}{2\sqrt{\pi(1+\sigma)}}
\Big( \cosh\big(\tfrac{2y\epsilon }{1+\sigma}\big) \big(\rho + \Sz \rho \Sz\big) - \sinh\big(\tfrac{2y\epsilon }{1+\sigma}\big) \big( \Sz \rho +\rho \Sz\big) + e^{-\frac{\sigma \epsilon^2}{1+\sigma}} \big( \rho - \Sz \rho \Sz\big) \Big)
.
\end{multline*}
and
$$
\Tr{\KK_y(\rho)} = \frac{e^{-\frac{y^2+\epsilon^2}{1+\sigma} }}{2\sqrt{\pi(1+\sigma)}}
\left(  \cosh\big(\tfrac{2y\epsilon }{1+\sigma}\big) - \sinh\big(\tfrac{2y\epsilon }{1+\sigma}\big) \Tr{\Sz\rho}  \right)
$$
one gets
\begin{multline*}
\frac{\KK_y(\rho)}{\Tr{\KK_y(\rho)}} =
\frac{
\left(1 + \frac{e^{-\tfrac{\sigma \epsilon^2}{1+\sigma}}}{ \cosh\big(\tfrac{2y\epsilon }{1+\sigma}\big)}\right) \rho
+
\left(1 + \frac{e^{-\tfrac{\sigma \epsilon^2}{1+\sigma}}}{ \cosh\big(\tfrac{2y\epsilon }{1+\sigma}\big)}\right) \Sz\rho \Sz
- \tanh\big(\tfrac{2y\epsilon }{1+\sigma}\big)  (\Sz\rho + \rho \Sz)
 }
{2\left(1 - \tanh\big(\tfrac{2y\epsilon }{1+\sigma}\big) \Tr{\Sz\rho}\right)}
.
\end{multline*}
Up to third order terms versus $\epsilon y$, one has then
$$
\frac{\KK_y(\rho)}{\Tr{\KK_y(\rho)}}
= \rho + \left(\tfrac{(\epsilon y)^2}{(1+\sigma)^2} + \tfrac{\sigma \epsilon^2}{2(1+\sigma)} \right) (\Sz\rho \Sz - \rho)  - \left( \tfrac{\epsilon y}{1+\sigma} + \tfrac{2(\epsilon y)^2}{(1+\sigma)^2}\Tr{\Sz\rho}\right) \big(\Sz\rho + \rho\Sz - 2 \Tr{\Sz\rho} \rho  \big)
.
$$
Replacing $\epsilon^2 y^2$ by its average independent of $\rho$ one gets
$$
\frac{\KK_y(\rho)}{\Tr{\KK_y(\rho)}}
= \rho + \tfrac{\epsilon^2}{2}  (\Sz\rho \Sz - \rho)  - \left( \tfrac{\epsilon y}{1+\sigma} + \tfrac{\epsilon^2}{1+\sigma}\Tr{\Sz\rho}\right) \big(\Sz\rho + \rho\Sz - 2 \Tr{\Sz\rho} \rho  \big)
.
$$
Set $\epsilon^2=2dt$ and $\epsilon y = -2 \Tr{\Sz\rho}dt - \sqrt{1+\sigma}~ dW$.
Since  by construction
$$
\EE{\epsilon y}{\rho}= -\epsilon^2 \Tr{\Sz \rho} \text{ and } \EE{(\epsilon y)^2}{\rho}= \epsilon^2/(1+\sigma) + \epsilon^4
$$   one has
$\EE{dW}{\rho} = 0$ and $\EE{dW^2}{\rho}=dt$. Thus for $dt$ very small, we recover the following SME  with detection efficiency $\eta = \frac{1}{1+\sigma}$
$$
\rho_{t+dt} = \rho_t + dt  \Big(\Sz \rho_t \Sz -\rho\Big)+\sqrt{\eta}\Big(\Sz \rho_t + \rho_t \Sz - 2\Tr{\Sz\rho_t} \rho\Big)  dW_t
$$
with $dy_t= \sqrt{\eta} \Tr{\Sz\rho_t+\rho_t\Sz} + dW_t $  corresponding to $- \epsilon y /\sqrt{1+\sigma}$ and $dW_t$ a Wiener process satisfying Ito rules  $dW_t^2=dt$.

Convergence towards either $\ket{g}$ or $\ket{e}$  is  based on the following Lyapunov fonction   $V(\rho)= \sqrt{1-\Tr{\Sz \rho}^2}$. According to Ito rules, one has
$$
dV= - \frac{z dz}{\sqrt{1-z^2}} -\frac{dz^2}{2(1-z^2)^{3/2} } = - \frac{z dz}{\sqrt{1-z^2}} -2\eta^2 V dt
$$
where $z=\Tr{\Sz\rho}$,  $dz=2\eta(1-z^2) dW $ and $dz^2=4 \eta^2 (1-z^2)^2 dt$.
Since $\EE{dz}{z}=0$,  $\bar V_t=\EE{V(z_t)}{z_0}$ converges exponentially to $0$ since  governed by the linear differential equation
$$
\dotex \bar V_t = - 2 \eta^2 \bar V_t, \quad \bar V_0= V(z_0).
$$
For more  general and  precise results on  diffusive SME corresponding to QND measurements and measurement-based feedback issues  see~\cite{BauerIHP2013,LiangAminiSiam2019,CardonaAUTO2020}.

\subsection{Diffusive SME}
As studied in~\cite{BarchielliGregorattiBook}, the general form of diffusive SME admits the  following Ito formulation:
\begin{align*}
d\rho_t &= \left(-i[\bH,\rho_t]+\sum_{\nu}\bL_{\nu}\rho_t\bL_{\nu}^\dag -\frac{1}{2}(\bL_{\nu}^\dag\bL_{\nu}\rho_t+\rho_t\bL_{\nu}^\dag\bL_{\nu})   \right)dt\\
&+ \sum_{\nu}\sqrt{\eta_{\nu}}\left(\bL_{\nu}\rho_t+\rho_t\bL_{\nu}^\dag-\Tr{(\bL_{\nu}+\bL_{\nu}^\dag)\rho_t}\rho_t\right)dW_{\nu,t},
\\
dy_{\nu,t} &= \sqrt{\eta_\nu} \Tr{\bL_\nu \rho_t + \rho_t \bL_{\nu}^\dag} dt + dW_{\nu,t}
\end{align*}
with efficiencies $\eta_\nu \in[0,1]$ and $dW_{\nu,t}$ being independent Wiener processes.
Here the Hilbert space $\mathcal{H}$ is arbitrary, $\bH$ is Hermitian and $\bL_\nu$ are arbitrary operators of $\mathcal{H}$ not necessarily Hermitian. Each  label $\mu$ such that $\eta_\mu=0$ corresponds here to a  decoherence channel that can be seen as  an unread measurement performed by a  sub-system belonging to  the environment, see~\cite[chapter 4]{haroche-raimondBook06}.

 With Ito rules, this SME admits also the following equivalent formulation:
$$
\rho_{t+dt}=\frac{\bM_{dy_t}\rho_t\bM_{dy_t}^\dag+\sum_\nu (1-\eta_\nu)\bL_{\nu}\rho_t\bL_\nu^\dag dt}{\Tr{\bM_{dy_t}\rho_t\bM_{dy_t}^\dag+\sum_\nu (1-\eta_\nu)\bL_{\nu}\rho_t\bL_\nu^\dag dt}}
$$
with
$$
\bM_{dy_t}=\bI+\left(-i\bH-\frac{1}{2}\sum_\nu\bL_{\nu}^\dag\bL_\nu \right)dt+\sum_{\nu}\sqrt{\eta_\nu}dy_{\nu,t}\bL_\nu.
$$
Moreover $dy_{\nu,t}=s_{\nu,t} \sqrt{dt} $ follows the following probability density  knowing $\rho_t$:
$$
\hspace{-1em}\PP\Big(\left(s_{\nu,t}\in[s_\nu,s_\nu+ds_\nu]\right)_\nu~|~\rho_t\Big)=
\Tr{\bM_{s\sqrt{dt}}~\rho_t\bM_{s\sqrt{dt}}^\dag+\sum_\nu(1-\eta_\nu)\bL_\nu\rho_t\bL_{\nu}^\dag dt } \prod_\nu \tfrac{e^{-\frac{s_\nu^2}{2}}ds_\nu}{\sqrt{2\pi}}
$$
that remains  a linear  function of $\rho$, as  imposed  by the quantum  measurement  law.

 In finite dimension $N$ , this formulation implies directly that any diffusive SME  admits  a unique  solution remaining for all $t\geq 0$ in  $\{\rho\in\CC^{N\times N}~:~\rho=\rho^\dag,~\rho\geq 0,~\Tr{\rho}=1\}$.

\subsection{Kraus maps and  numerical schemes for diffusive SME}
From the above formulation, one can construct a linear,  positivity and trace preserving numerical integration scheme for such diffusive SME (see~\cite[appendix B]{JordanChantasriRouchonEtAl2016}):
\begin{align*}
d\rho_t &= \left(-i[\bH,\rho_t]+\sum_{\nu}\bL_{\nu}\rho_t\bL_{\nu}^\dag -\frac{1}{2}(\bL_{\nu}^\dag\bL_{\nu}\rho_t+\rho_t\bL_{\nu}^\dag\bL_{\nu})   \right)dt\\
&+ \sum_{\nu}\sqrt{\eta_{\nu}}\left(\bL_{\nu}\rho_t+\rho_t\bL_{\nu}^\dag-\Tr{(\bL_{\nu}+\bL_{\nu}^\dag)\rho_t}\rho_t\right)dW_{\nu,t},
\\
dy_{\nu,t} &= \sqrt{\eta_\nu} \Tr{\bL_\nu \rho_t + \rho_t \bL_{\nu}^\dag} dt + dW_{\nu,t}
\end{align*}
With $$
\bM_0=\bI+\big(-i\bH-\tfrac{1}{2}\sum_\nu\bL_{\nu}^\dag\bL_\nu \big)dt, \quad
\bS= \bM_0^\dag \bM_0+ \left( \sum_\nu  \bL^\dag_\nu \bL_\nu\right) dt
$$
set
$$
\btM_0= \bM_0 \bS^{-1/2},\quad \btL_\nu=\bL_\nu \bS^{-1/2}
$$
Sampling of   $dy_{\nu,t}=s_{\nu,t} \sqrt{dt} $ according to the following   probability law:
$$
\hspace{-1em}\PP\Big(\left(s_{\nu,t}\in[s_\nu,s_\nu+ds_\nu]\right)_\nu~|~\rho_t\Big)=
\Tr{\btM_{s\sqrt{dt}}\rho_t\btM_{s\sqrt{dt}}^\dag+\sum_\nu(1-\eta_\nu)\btL_\nu\rho_t\btL_{\nu}^\dag dt } \prod_\nu \tfrac{e^{-\frac{s_\nu^2}{2}}ds_\nu}{\sqrt{2\pi}}.
$$
where
$$
\btM_{dy_t}=\btM_0+\sum_{\nu}\sqrt{\eta_\nu}dy_{\nu,t}\btL_\nu.
$$
The  update  $\rho_{t+dt}$ is then  given by the following exact Kraus-map formulation:
$$
\rho_{t+dt}=\frac{\btM_{dy_t}\rho_t\btM_{dy_t}^\dag+\sum_\nu (1-\eta_\nu)\btL_{\nu}\rho_t\btL_\nu^\dag dt}
{\Tr{\btM_{dy_t}\rho_t\btM_{dy_t}^\dag+\sum_\nu (1-\eta_\nu)\btL_{\nu}\rho_t\btL_\nu^\dag dt}}
.
$$
Notice that the operators $\btM_{dy_t}$ and $\btL_{\nu}$ are   bounded operators even if $\bH$ and $\bL_\nu$ are unbounded.

 One can also use the following splitting  scheme when the unitary operator $e^{-\frac{i dt}{2} \bH}$  is numerically available and where  in the above calculations $\bM_0$ is reduced  to
 $ \bI-\tfrac{dt}{2}\sum_\nu\bL_{\nu}^\dag\bL_\nu $:
 $$
\rho_{t+dt}=e^{-\frac{i dt}{2} \bH} \frac{\btM_{dy_t}e^{-\frac{i dt}{2} \bH}\rho_t e^{\frac{i dt}{2} \bH}\btM_{dy_t}^\dag+\sum_\nu (1-\eta_\nu)\btL_{\nu}e^{-\frac{i dt}{2} \bH}\rho_t e^{\frac{i dt}{2} \bH}\btL_\nu^\dag dt}
{\Tr{\btM_{dy_t}e^{-\frac{i dt}{2} \bH}\rho_t e^{\frac{i dt}{2} \bH}\btM_{dy_t}^\dag+\sum_\nu (1-\eta_\nu)\btL_{\nu}e^{-\frac{i dt}{2} \bH}\rho_t e^{\frac{i dt}{2} \bH}\btL_\nu^\dag dt}}
e^{\frac{i dt}{2} \bH}
.
$$

\subsection{Qubits measured by photons (resonant interaction)}
The qubit wave function is denoted by $\ket{\psi}$. The photons,  before interacting with  the qubit,  are in the   vacuum   state $\ket{0}$.  The interaction is resonant  according to~\eqref{eq:resonantPropagator}.
After the interaction and just before the measurement performed on the photons, the composite qubit/photon wave function $\ket{\Psi}$ reads:
\begin{multline*}
\Bigg(\ketbra{g}{g} \cos( \theta \sqrt{\bn})+ \ketbra{e}{e} \cos( \theta \sqrt{\bn+\bI})
\\
+\ketbra{g}{e} \frac{\sin( \theta \sqrt{\bn})}{ \sqrt{\bn}} \ba^\dag
-\ketbra{e}{g} \ba  \frac{\sin( \theta \sqrt{\bn})}{ \sqrt{\bn}}\Bigg) \ket{\psi}\ket{0}
\\
= \big( \braket{g}{\psi} \ket{g}+  \cos\theta \braket{e}{\psi} \ket{e}\big) \ket{0}
+ \sin\theta \braket{e}{\psi} \ket{g} \ket{1}
.
\end{multline*}
The Markov process  with measurement   observable $\bn=\sum_{n\geq 0} n \ketbra{n}{n}$ and outcome $y\in\{0, 1\}$ reads (density operator formulation)
$$
\rho_{k+1} = \left\{
                     \begin{array}{ll}
                      \frac{\bM_0 \rho_k \bM_0^\dag}{\Tr{\bM_0 \rho_k \bM_0^\dag}} \, & \hbox{if $y_k=0$ with probability $\Tr{\bM_0 \rho_k \bM_0^\dag}$ ;} \\
                      \frac{\bM_1 \rho_k \bM_1^\dag}{\Tr{\bM_1 \rho_k \bM_1^\dag}} \, & \hbox{if $y_k=1$ with probability $\Tr{\bM_1 \rho_k \bM_1^\dag}$ ;}
                     \end{array}
                   \right.
$$
with measurement Kraus operators $\bM_0=\ketbra{g}{g}+\cos\theta \ketbra{e}{e} $ and $\bM_1=\sin\theta \ketbra{g}{e}$.
Notice that  $\bM_0^\dag \bM_0 + \bM_1^\dag \bM_1=\bI$.

Almost convergence analysis when $\cos^2(\theta) <1$  towards  $\ket{g}$ can be seen via  the Lyapunov function (super martingale)
$$
V(\rho)=\Tr{\ketbra{e}{e} \rho}
$$
since
$$
\EE{V(\rho_{k+1})}{\rho_k} = \cos^2\theta ~ V(\rho_{k})
.
$$

\subsection{Towards  jump SME}

Since in the above Markov process  $\Tr{\bM_0\rho\bM_0^\dag}= 1 - \sin^2\theta \Tr{\Sm \rho\Sp}$ and  $\Tr{\bM_1\rho\bM_1^\dag}= \sin^2\theta \Tr{\Sm \rho\Sp }$,
 one gets with $\sin^2\theta = dt$ and $y$ being denoted by $dN$,   an SME driven by Poisson   process $dN_t\in\{0,1\}$  of  expectation value  $\Tr{\Sm \rho_t \Sp } dt$ knowing $\rho_t$:
\begin{multline*}
 d\rho_t =\left(\Sm\rho_t \Sp - \tfrac{1}{2}(\Sp \Sm\rho_t +\rho_t \Sp\Sm) \right)\,dt
\\ +
   \left(\frac{ \Sm\rho_t \Sp}{\Tr{\Sm\rho_t \Sp}} -\rho_t\right)
      \left({{dN_t }}-\Big( \Tr{\Sm\rho_t \Sp}\Big)\,dt\right)
.
\end{multline*}
At each time-step, one has the following choice:
\begin{itemize}
  \item with probabilty  $ 1-  \Tr{\Sm\rho_t \Sp}\,dt$,  $dN_t=N_{t+dt}-N_t =0$ and
$$
\rho_{t+dt}= \frac{
 \bM_{0}\rho_t \bM_{0}^\dag
                          }{
                          \Tr{
 \bM_{0}\rho_t \bM_{0}^\dag
                          }}
$$
with $\bM_{0} = \bI - \tfrac{dt}{2} \Sp \Sm $.

\item with probability $\Tr{\Sm\rho_t \Sp}\,dt$,   $ dN_{t}=N_{t+dt}-N_{t}=1$ and
$$
\rho_{t+dt}= \frac{
 \bM_{1}\rho_t \bM_{1}^\dag
                          }{
                          \Tr{
 \bM_{1}\rho_t \bM_{1}^\dag
                          }}
$$
with $\bM_1= \sqrt{dt} ~\Sm$.
\end{itemize}

To take into account shot noise  of rate $\bar\theta\geq 0 $ and detection efficiency $\bar\eta \in[0,1]$, consider the following left stochastic  matrix
$$
\left(
  \begin{array}{cccc}
   1- \bar\theta dt  & 1- \bar \eta
 \\
  \bar\theta dt & \bar \eta
  \end{array}
\right)
$$
where $\bar\theta dt$ is the probability to detect $y=1$, knowing that the true outcome is $0$ (fault detection associated to shot noise) and where $\bar\eta$ is the probability to detect $y=1$ knowing that the true outcome is $1$ (detection efficiency).
Then the above stochastic master equation becomes
\begin{multline*}
  d\rho_t =\left(\Sm\rho_t \Sp - \tfrac{1}{2}(\Sp \Sm\rho_t +\rho_t \Sp\Sm) \right)\,dt
\\+
   \left(\frac{ \bar\theta \rho_t +\bar \eta \Sm\rho_t \Sp}{\Tr{\bar\theta \rho_t  + \bar \eta \Sm\rho_t \Sp}} -\rho_t\right)
      \left({{dN_t }}-\Big(\bar\theta +\bar \eta\Tr{\Sm\rho_t \Sp}\Big)\,dt\right)
.
\end{multline*}
At each time-step, one has the following recipe
\begin{itemize}
  \item  $dN_t=N_{t+dt}-N_t =0$ and
\begin{multline*}
 \rho_{t+dt}= \frac{
 (1-\bar\theta dt) \bM_{0}\rho_t \bM_{0}^\dag + (1-\bar\eta)  \bM_{1}\rho_t \bM_{1}^\dag
                          }{
                          \Tr{
  (1-\bar\theta dt) \bM_{0}\rho_t \bM_{0}^\dag + (1-\bar\eta)  \bM_{1}\rho_t \bM_{1}^\dag
                          }}
                          \\
                          = \frac{
 \bM_{ { {0}}}\rho_t \bM_{ { {0}}}^\dag + (1-\bar{\eta} )\bM_{1}\rho_t \bM_{1}^\dag
                          }{
                          \Tr{
 \bM_{ { {0}}}\rho_t \bM_{ { {0}}}^\dag + (1-\bar{\eta}) \bM_{1}\rho_t \bM_{1}^\dag
                          }} + O(dt^2).
\end{multline*}
with probability
$$
1-\Big(\bar\theta +\bar\eta\Tr{\Sm\rho_t \Sp}\Big) dt = \Tr{(1-\bar\theta dt) \bM_{0}\rho_t \bM_{0}^\dag + (1-\bar\eta)  \bM_{1}\rho_t \bM_{1}^\dag}+ O(dt^2)
$$
and  where  $\bM_{0} = \bI - \tfrac{dt}{2} \Sp \Sm $ and  $\bM_1= \sqrt{dt} ~\Sm$.

\item   $ dN_{t}=N_{t+dt}-N_{t}=1$ and
\begin{multline*}
 \rho_{t+dt}= \frac{
\bar\theta\,dt\,  \bM_{0}\rho_t \bM_{0}^\dag + \bar\eta  \bM_{1}\rho_t \bM_{1}^\dag
                        }{
                          \Tr{
\bar\theta\,dt\,  \bM_{0}\rho_t \bM_{0}^\dag + \bar\eta  \bM_{1}\rho_t \bM_{1}^\dag
                          }}
                          =  \frac{ \bar{\theta}\rho_t+\bar{\eta} \Sm\rho_t \Sp}{\bar{\theta}+\bar{\eta}\Tr{\Sm\rho_t \Sp}} + O(dt)
\end{multline*}
with probability
$$
(\bar\theta + \bar\eta \Tr{\Sm\rho_t \Sp}\Big)dt
=  \Tr{\bar\theta\,dt\,  \bM_{0}\rho_t \bM_{0}^\dag +\bar \eta  \bM_{1}\rho_t \bM_{1}^\dag} + O(dt^2)
$$
\end{itemize}

\subsection{Jump SME in continuous-time}

The above  computations with $dt$ very small emphasize the following general structure of a Jump SME in continuous time.
With the counting  process ${{N_t}}$ having  increment  expectation value knowing $\rho_t$  given by
$
{{ \langle dN_t \rangle}}= \Big( \bar{\theta}+\bar{\eta} \Tr{V \rho_t V^\dag}\Big)\,dt,
$  and detection  imperfections modeled by $\bar{\theta}\geq 0$ (shot-noise rate)  and $\bar{\eta} \in [0,1]$ (detection efficiency),
the quantum state $\rho_t$  is usually mixed and obeys to
\begin{multline*}
d\rho_t =\left(-i[\bH,\rho_t] +  \bV\rho_t \bV^\dag - \tfrac{1}{2}(\bV^\dag \bV\rho_t +\rho_t \bV^\dag \bV) \right)\,dt
\\
+
   \left(\frac{\bar{\theta}\rho_t+\bar{\eta} \bV\rho_t \bV^\dag}{\bar{\theta}+\bar{\eta}\Tr{\bV\rho_t \bV^\dag}} -\rho_t\right)
      \left( { {dN_{t} }}-\Big( \bar{\theta} +\bar{\eta} \Tr{\bV\rho_t \bV^\dag}\Big)\,dt\right)
      .
\end{multline*}
Here $\bH$ and $\bV$ are operators on an underlying  Hilbert space $\mathcal{H}$, $\bH$ being Hermitian.
At each time-step between $t$  and $t+dt$, one has
\begin{itemize}
  \item  $ { {dN_{t} =0}}$  with probability $ 1- \Big( \bar{\theta}+\bar{\eta} \Tr{\bV \rho_t \bV^\dag}\Big)\,dt$
$$
\rho_{t+dt}= \frac{
 \bM_{ { {0}}}\rho_t \bM_{ { {0}}}^\dag + (1-\bar{\eta} ) \bV \rho_t \bV^\dag dt
                          }{
                          \Tr{
 \bM_{ { {0}}}\rho_t \bM_{ { {0}}}^\dag + (1-\bar{\eta}) \bV\rho_t \bV^\dag dt
                          }}
$$
where $\bM_{ { {0}}} = I - \left(iH + \tfrac{1}{2} \bV^\dag \bV\right) dt $.

\item  $ { {dN_{t} =1}}$ with probability  $\Big( \bar{\theta}+\bar{\eta} \Tr{\bV \rho_t \bV^\dag}\Big)\,dt$,
$$
\rho_{t+dt}= \frac{ \bar{\theta}\rho_t+\bar{\eta} \bV\rho_t \bV^\dag}{\bar{\theta}+\bar{\eta}\Tr{\bV\rho_t \bV^\dag}}
                      .
$$
These SME have been introduced in the Physics literature in~\cite{DalibCM1992PRL,GardiPZ1992PRA}.
\end{itemize}

\subsection{General mixed diffusive/jump SME}
One can combine in  a single SME  Wiener  and Poisson  noises induced by diffusive and counting measurements.
The quantum state $\rho_t$, usually mixed,   obeys to
\begin{multline*}
\hspace{-2em} d\rho_t =\left(-i[\bH,\rho_t] +  \bL \rho_t \bL^\dag - \tfrac{1}{2} (\bL^\dag \bL\rho_t+\rho_t \bL^\dag \bL)+ \bV\rho_t \bV^\dag - \tfrac{1}{2}(\bV^\dag \bV\rho_t +\rho_t \bV^\dag \bV) \right)\,dt
\\
+  \sqrt{\eta} \bigg(\bL\rho_t+\rho_t \bL^\dag-\Tr{(\bL+\bL^\dag)\rho_t}\rho_t\bigg) { {dW_{t}}}
\\
 +  \left(\frac{\bar{\theta}\rho_t+\bar{\eta} \bV\rho_t \bV^\dag}{\bar{\theta}+\bar{\eta}\Tr{\bV\rho_t \bV^\dag}} -\rho_t\right)
      \left( { {dN_{t} }}-\Big( \bar{\theta} +\bar{\eta} \Tr{\bV\rho_t \bV^\dag}\Big)\,dt\right)
\end{multline*}
\\[1.em] With $ { {dy_{t}}}=  \sqrt{\eta} \Tr{(\bL+\bL^\dag)\,\rho_t}\,dt +  { {dW_{t}}}$ and
$ { {dN_{t} =0}}$ with probability  $ 1- \Big( \bar{\theta}+\bar{\eta} \Tr{\bV \rho_t \bV^\dag}\Big)\,dt$.
The Kraus-map equivalent formulation reads:
\begin{itemize}
 \item
 for $dN_t=0$ of probability  $1-\Big( \bar{\theta}+\bar{\eta} \Tr{\bV \rho_t \bV^\dag}\Big)\,dt$
$$
\rho_{t+dt}= \frac{
 \bM_{ { {dy_{t}}}}\rho_t \bM_{ { {dy_{t}}}}^\dag + (1-\eta ) \bL\rho_t \bL^\dag dt+  (1-\bar{\eta} ) \bV \rho_t \bV^\dag dt
                          }{
                          \Tr{
 \bM_{ { {dy_{t}}}}\rho_t \bM_{ { {dy_{t}}}}^\dag + (1-\eta ) \bL\rho_t \bL^\dag dt+ (1-\bar{\eta}) \bV\rho_t \bV^\dag dt
                          }}
$$
with $\bM_{ { {dy_{t}}}} = I - \left(i\bH +\tfrac{1}{2} \bL^\dag \bL + \tfrac{1}{2} \bV^\dag \bV\right) dt + \sqrt{\eta}  { {dy_{t}}} \bL$.

\item  for $dN_t=1$ of probability  $\Big( \bar{\theta}+\bar{\eta} \Tr{\bV \rho_t \bV^\dag}\Big)\,dt$:
$$
\rho_{t+dt}= \frac{
 \bM_{ { {dy_{t}}}}\tilde\rho_t \bM_{ { {dy_{t}}}}^\dag + (1-\eta ) \bL\tilde\rho_t \bL^\dag dt+  (1-\bar{\eta} ) \bV \tilde\rho_t \bV^\dag dt
                          }{
                          \Tr{
 \bM_{ { {dy_{t}}}}\tilde\rho_t \bM_{ { {dy_{t}}}}^\dag + (1-\eta ) \bL\tilde\rho_t \bL^\dag dt+ (1-\bar{\eta}) \bV\tilde\rho_t \bV^\dag dt
                          }}
                         \text{ with }  \tilde\rho_t=\displaystyle\frac{ \bar{\theta}\rho_t+\bar{\eta} \bV\rho_t \bV^\dag}{\bar{\theta}+\bar{\eta}\Tr{\bV\rho_t \bV^\dag}}
$$
\end{itemize}

More generally, one can consider  several independent Wiener and Poisson  processes. The corresponding  SME reads then
{\small
\begin{multline*}
d\rho_t =\left(-i[\bH,\rho_t] +  \sum_\nu \bL_{\nu} \rho_t \bL_{\nu}^\dag - \tfrac{1}{2}(\bL_{\nu}^\dag \bL_{\nu}\rho_t+\rho_t \bL_{\nu}^\dag \bL_{\nu})+ \sum_\mu \bV_{\mu}\rho_t \bV_{\mu}^\dag - \tfrac{1}{2}(\bV_{\mu}^\dag \bV_{\mu}\rho_t +\rho_t \bV_{\mu}^\dag \bV_{\mu}) \right)\,dt
\\
+ \sum_\nu  \sqrt{\eta_\nu} \bigg(\bL_{\nu}\rho_t+\rho_t \bL_{\nu}^\dag-\Tr{(\bL_{\nu}+\bL_{\nu}^\dag)\rho_t}\rho_t\bigg) { {dW_{\nu,t}}}
\\
 +  \sum_\mu \left(\frac{\bar{\theta}_\mu\rho_t+\sum_{\mu'} \bar{\eta}_{\mu,\mu'} \bV_{\mu'}\rho_t \bV_{\mu'}^\dag}{\bar{\theta}_\mu +
   \sum_{\mu'}\bar{\eta}_{\mu,\mu'}\Tr{\bV_{\mu'}\rho_t \bV_{\mu'}^\dag}} -\rho_t\right)
      \left( { {dN_{\mu,t} }}-\Big( \bar{\theta}_\mu + \sum_{\mu'}\bar{\eta}_{\mu,\mu'} \Tr{\bV_{\mu'}\rho_t \bV_{\mu'}^\dag}\Big)\,dt\right)
\end{multline*}
}
where  $\eta_\nu \in[0,1]$, $\bar{\theta}_\mu, \bar{\eta}_{\mu,\mu'}\geq 0$ with $\bar{\eta}_{\mu'}=\sum_{\mu} \bar{\eta}_{\mu,\mu'} \leq 1$ are parameters modelling measurements  imperfections.

The equivalent Kraus-map formulation is the following
\begin{itemize}

\item When  $\forall \mu,~ { {dN_{\mu,t} =0}}$ (probability $1-\sum_\mu\Big( \bar{\theta}_\mu+\bar{\eta}_\mu \Tr{\bV_\mu \rho_t \bV_\mu^\dag}\Big)\,dt$) we have
$$
\rho_{t+dt}= \frac{
 \bM_{ { {dy_{t}}}}\rho_t \bM_{ { {dy_{t}}}}^\dag + \sum_\nu (1-\eta_\nu ) \bL_{\nu}\rho_t \bL_{\nu}^\dag dt+  \sum_\mu (1-\bar{\eta}_\mu ) \bV_{\mu} \rho_t \bV_{\mu}^\dag dt
                          }{
                          \Tr{
 \bM_{ { {dy_{t}}}}\rho_t \bM_{ { {dy_{t}}}}^\dag + \sum_\nu (1-\eta_\nu ) \bL_{\nu}\rho_t \bL_{\nu}^\dag dt+ \sum_\mu (1-\bar{\eta}_\mu) \bV_{\mu}\rho_t \bV_{\mu}^\dag dt
                          }}
$$
with $\bM_{ { {dy_{t}}}} = I - \left(iH +\tfrac{1}{2} \sum_\nu \bL_{\nu}^\dag \bL_{\nu} + \tfrac{1}{2} \sum_\mu\ \bV_{\mu}^\dag \bV_{\mu}\right) dt + \sum_\nu \sqrt{\eta_\nu}  { {dy_{\nu t}}} \bL_{\nu}$ and
where  $ { {dy_{\nu,t}}}=  \sqrt{\eta_\nu} \Tr{(\bL_\nu+\bL_\nu^\dag)\,\rho_t}\,dt +  { {dW_{\nu,t}}}$.

\item
If, for some $\mu$,  $  dN_{\mu,t}=1$ (probability $\Big( \bar{\theta}_\mu+\sum_{\mu'}\bar{\eta}_{\mu,\mu'} \Tr{\bV_{\mu'} \rho_t \bV_{\mu'}^\dag}\Big)\,dt$) we have  a similar transition rule
$$
\rho_{t+dt}= \frac{
 \bM_{ { {dy_{t}}}}\tilde\rho_t \bM_{ { {dy_{t}}}}^\dag + \sum_\nu (1-\eta_\nu ) \bL_{\nu}\tilde\rho_t \bL_{\nu}^\dag dt+  \sum_{\mu'} (1-\bar{\eta}_{\mu'} ) \bV_{\mu'} \tilde\rho_t \bV_{\mu'}^\dag dt
                          }{
                          \Tr{
 \bM_{ { {dy_{t}}}}\tilde\rho_t \bM_{ { {dy_{t}}}}^\dag + \sum_\nu (1-\eta_\nu ) \bL_{\nu}\tilde\rho_t \bL_{\nu}^\dag dt+ \sum_{\mu'} (1-\bar{\eta}_{\mu'}) \bV_{\mu'}\tilde\rho_t \bV_{\mu'}^\dag dt
                          }}
 $$
 with
 $
 \tilde\rho_{t}=\tfrac{ \bar{\theta}_\mu\rho_t+\sum_{\mu'}\bar{\eta}_{\mu,\mu'} \bV_{\mu'}\rho_t \bV_{\mu'}^\dag}{\bar{\theta}_\mu+\sum_{\mu'} \bar{\eta}_{\mu,\mu'}\Tr{\bV_{\mu'}\rho_t \bV_{\mu'}^\dag}}
 .
$
\end{itemize}

\section{Conclusion} \label{sec:conclusion}

These SME  driven by diffusive measurements or counting measurements are now the object of numerous control-theoretical and mathematical investigations that can be divided into two main issues.
The first issues are related to feedback  stabilization of a target quantum state (quantum state preparation)   or of  quantum subspace as in quantum error correction. One can distinguish several kinds of quantum feedback:
\begin{itemize}
  \item Markovian feedback~\cite{wiseman-milburnBook} which is in fact a static output feedback  usually used in discrete-time  quantum error correction. Its main interest relies on the closed-loop ensemble average dynamics which is a linear quantum channel for which several stability properties are available (see, e.g.,\cite{nielsen-chang-book,petz:LAA1996,SepulSR2010}).

  \item measurement-based feedback where the control-loop is still achieved by a classical  controller taking into account the past measurement outcomes. Quantum-state feedback is  such typical feedback where the quantum state is estimated via a quantum filter (see, e.g.,\cite{ahn-et-alPRA02,sayrin-et-al:nature2011,AminiSDSMR2013A,LiangAminiSiam2019,CardonaAUTO2020}).

  \item Coherent feedback where the controller is a quantum dissipative system~\cite{GoughJ2009ACITo}. It has its origin in optical pumping and coherent population trapping~\cite{Kastl1967S,arimondo-96}. Such feedback structures are  now the object of active researches in the context of autonomous quantum error correction (see, e.g., \cite{VerstraeteWolfIgnacioCirac2009,sarlette-et-al:PRL2011, LeghtKVSDM2013PRL,MirrahimiCatComp2014,LescanneZaki2019}).
\end{itemize}

The  second issues are related to filtering and estimation. They are  closely related to   quantum-state or quantum-process tomography:
\begin{itemize}

  \item Quantum filtering that can be seen as the quantum analogue of  state asymptotic observers. It   has its origin in the seminal work of   \cite{Belavkin1992}. It can be shown that quantum filtering is always a stable process in average (see~\cite{Rouch2011ACITo,AminiPR2014}).  Characterization of asymptotic almost-sure convergence is an open-problem with recent progresses  (see, e.g.,  \cite{van-handel:proba2009,Amini_2021}).

  \item Estimation of  a quantum state or  classical parameters  estimation  based on repeated measurements including imperfection and decoherence during the measurement process relies on quantum  SME (see, e.g., \cite{SixPRA2016,TilloyPRA2018})

\end{itemize}

\paragraph{Acknowledgment} The author thanks Claude Le Bris,  Philippe Campagne-Ibarcq, Zaki Leghtas, Mazyar Mirrahimi, Alain Sarlette and Antoine Tilloy for many interesting discussions on SME modeling and numerics  for  open quantum systems.

This project has received funding from the European Research Council (ERC) under the European Union’s Horizon 2020 research and innovation programme (grant agreement No. [884762]).


\appendix
\section{Notations used for qubits and photons} \label{ap:qubitphoton}
\begin{enumerate}
\item  The qubit with a two-dimensional Hilbert space:
   \begin{itemize}
    \item Hilbert  space: \mbox{${\mathcal{H}}=\CC^2=\Big\{\psi_g\ket{{g}}+\psi_e\ket{{e}},\; \psi_g,\psi_e\in \CC\Big\}$} with ortho-normal basis $\ket{g}$ and $\ket{e}$ (Dirac notations).
    \item  Quantum state space: $\mathcal{D}=\{\rho\in\mathcal{L}(\mathcal{H}), \rho^\dag=\rho, \Tr{\rho}=1,\rho\ge 0\}\;.$
    \item Pauli operators and commutations:
    \\
    $\Sm=\ket{{g}}\bra{{e}}$, $\Sp=\Sm^\dag=\ket{{e}}\bra{{g}}$
    \\
    ${\Sx}=\Sm+\Sp=\ket{{g}}\bra{{e}}+ \ket{{e}}\bra{{g}}$;
    \\
    $\Sy=i \Sm-i \Sp = i\ket{{g}}\bra{{e}}-i\ket{{e}}\bra{{g}}$;
    \\
    ${\Sz}=\Sp\Sm - \Sm \Sp = \ketbra{e}{e}-\ketbra{g}{g}$;
    \\
   $\Sx^2=\bI$,  $\Sx\Sy=i\Sz$,  $[\Sx,\Sy]=2i \Sz$, $ \ldots$

    \item Hamiltonian: $\bH={\omega_q \Sz}/2+{ {u_q}}{\Sx}$.

   \item Bloch sphere representation:\\
    \mbox{$\mathcal{D}=\Big\{\tfrac{1}{2} \big(\bI + x \Sx + y \Sy + z \Sz \big)~\big|~  (x,y,z)\in\RR^3,~x^2+y^2+z^2 \leq 1\Big\}$}
  \end{itemize}

\item  The photons of  the  quantum harmonic oscillator  with  an infinite dimensional Hilbert:
  \begin{itemize}
    \item Hilbert  space: $\mathcal{H}=\Big\{\sum_{n\ge 0} \psi_n \ket{n},\; (\psi_n)_{n\ge 0}\in l^2(\CC)\Big\}\equiv  L^2(\RR,\CC) $ with the infinite dimensional  orthonormal basis $\big( \ket{n}\big)_{n=0,1,2, \ldots}$.
    \item Quantum state space: $\DD=\{\rho\in\mathcal{L}(\mathcal{H}) \text{ trace class, }  \rho^\dag=\rho, \Tr{\rho}=1,\rho\ge 0\}\;$ corresponding to trace-class Hermitian operators on $\mathcal{H}$ with unit  trace.
    \item Operators and commutations:\\
     Annihilation and creation  operator:   ${\ba \ket{n}=\sqrt{n}~ {\ket{n- 1}}}$, $\ba^\dag \ket{n}=\sqrt{n+1}\ket{n+1}$;
    \\
    Number operator: $\bn=\ba^\dag \ba$,  $\bn\ket{n}=n\ket{n}$;
    \\
   ${ [\ba,\ba^\dag]=\bI}$,  $\ba f(\bn) = f(\bn+\bI) \ba $ for any function $f$;
   \\
    Coherent displacement unitary operator $\bD_\alpha=e^{\alpha\ba^\dag-\alpha^\dag \ba}$.
    \\
   Position $\bQ$ and momentum operators $\bP$: \\ $ \ba=\tfrac{\bQ+ i \bP}{\sqrt{2}}=\tfrac{1}{\sqrt{2}} \left(q+\frac{d}{dq}\right),~ [\bQ,\bP]=\imath \bI$.

    \item Hamiltonian: $\bH =\omega_c \ba^\dag\ba+{ {u_c}}(\ba+\ba^\dag)$. \\
    (associated classical dynamics:  $\frac{dq}{dt} = \omega_c p,~\frac{dp}{dt}=-\omega_c q - \sqrt{2} u_c$).
     \item  { Quasi-classical pure state $\equiv$ coherent state $\ket{\alpha}$}
  $\alpha\in\CC: ~\ket{\alpha}=\sum_{n\ge 0}{\left(e^{-|\alpha|^2/2}\frac{\alpha^n}{\sqrt{n!}}\right)\ket{n}}$;
    $ \ket{\alpha} \equiv  \tfrac{1 }{\pi^{1/4}}
e^{\imath\sqrt{2}q \Im\alpha } e^{-\frac{(q-\sqrt{2}\Re\alpha )^2}{2}}$;
 $\ba\ket{\alpha}=\alpha\ket{\alpha}$, $\bD_\alpha\ket{0}=\ket{\alpha}$.

  \end{itemize}

\end{enumerate}

\section{Three quantum rules} \label{ap:3Qrules}
This appendix is borrowed from~\cite{RouchonSeoul2014}
{ \em
\begin{enumerate}

  \item  The state of a quantum system is described either by the  wave function  $\ket{\psi}$ a vector of length one   belonging to some separable Hilbert space $\mathcal{H}$ of finite or infinite dimension, or,  more generally, by the density operator $\rho$ that is a non-negative  Hermitian operator on $\mathcal{H}$ with trace one. When the system can be described by a wave function $\ket{\psi}$ (pure state), the density operator  $\rho$ coincides with  the orthogonal projector on  the line spanned by $\ket{\psi}$ and   $\rho =\ket{\psi}\bra{\psi}$ with usual Dirac notations. In general the rank of $\rho$ exceeds one, the state is then mixed and cannot be described by a wave function.  When the system is closed, the  time evolution of $\ket{\psi}$ is  governed by the Schr\"{o}dinger equation (here $\hbar\equiv 1$)
      \begin{equation}\label{eq:Schrodinger}
        \dotex \ket\psi = - i\bH \ket\psi
      \end{equation}
      where $\bH$ is the system Hamiltonian, an Hermitian operator on $\mathcal{H}$  that could possibly depend on time $t$ via some time-varying parameters (classical control inputs).
       When the system is closed, the  evolution of $\rho$ is  governed by the Liouville/von-Neumann equation
      \begin{equation}\label{eq:Liouville}
        \dotex \rho = -i\big[\bH,\rho\big] =  -i \big(\bH\rho - \rho\bH\big)
        .
      \end{equation}

  \item Dissipation and irreversibility has its origin in the "collapse of the wave packet"  induced by the measurement. A measurement on the quantum system of state $\ket{\psi}$ or $\rho$  is associated  to   an observable  $\bO$, an Hermitian operator on $\mathcal{H}$,  with spectral decomposition $\sum_{\mu} \lambda_\mu \bP_\mu$:  $\bP_\mu$ is the orthogonal projector on the eigen-space associated to the eigen-value $\lambda_\mu$ ($\lambda_\mu \neq \lambda_{\mu'}$ for $\mu\neq\mu'$).  The measurement process attached to $\bO$ is assumed to be instantaneous and obeys to the following rules:
        \begin{itemize}
          \item the measurement outcome $\mu$ is obtained  with probability $\PP_\mu=\bra{\psi} \bP_\mu\ket\psi$ or $\PP_\mu=\Tr{\rho \bP_\mu}$,  depending on the state $\ket{\psi}$ or $\rho$ just before the measurement;
           \item just after the measurement process, the quantum state is changed to $\ket{\psi}_+$ or $\rho_+$ according to the mappings
               $$
  \ket\psi\mapsto\ket{\psi}_+ = \frac{\bP_\mu \ket{\psi}}{\sqrt{\bra\psi \bP_\mu\ket\psi}}\quad  \text{ or }\quad
  \rho\mapsto \rho_+ = \frac{\bP_\mu \rho \bP_\mu}{\Tr{\rho \bP_\mu}}
  $$
  where $\mu$  is  the observed  measurement outcome. These mappings  describe the measurement back-action and have  no classical counterpart.
        \end{itemize}

  \item Most systems are composite systems built with several sub-systems.  The quantum states of such composite systems live in the tensor product of the Hilbert spaces of each sub-system. This is a crucial difference with  classical composite systems where the state space is built  with Cartesian products. Such tensor products have important implications such   as  entanglement with   existence of non separable states. Consider a  bi-partite   system made of two sub-systems: the sub-system of interest $S$  with Hilbert space $\mathcal{H}_S$  and the measured sub-system $M$ with Hilbert space $\mathcal{H}_M$.  The quantum state of this bi-partite system $(S,M)$ lives in $\mathcal{H}=\mathcal{H}_S\otimes \mathcal{H}_M$. Its Hamiltonian $\bH$ is constructed with the Hamiltonians of the  sub-systems, $\bH_S$ and $\bH_M$, and an interaction  Hamiltonian $ \bH_{int}$ made of a sum of tensor products of operators (not necessarily Hermitian)  on $S$ and $M$:
      $$
      \bH=\bH_S\otimes\bI_M+\bH_{int}+\bI_S\otimes \bH_M
      $$
      with $\bI_S$ and $\bI_M$  identity operators on $\mathcal{H}_S$ and $\mathcal{H}_M$, respectively.
      The measurement operator $\bO=\bI_S \otimes \boldsymbol{O}_M $ is here  a simple tensor product of  identity on $\mathcal{H}_S$ and the Hermitian operator $\bO_M$ on $\mathcal{H}_M$, since only system $M$ is directly measured. Its spectrum is  degenerate: the multiplicities of the  eigenvalues are  necessarily greater or equal to the dimension of $\mathcal{H}_S$.

\end{enumerate}

}

\end{document}